\documentclass[sn-mathphys,iicol]{sn-jnl}

\usepackage{xspace}
\usepackage{caption}
\usepackage{subcaption}
\usepackage{textgreek}
\usepackage{siunitx}
\usepackage{dblfloatfix}    
\usepackage{cleveref}
\usepackage{placeins}

\usepackage{graphicx} 
\usepackage{circledsteps}
\usepackage{makecell}

\usepackage[normalem]{ulem}
\newcommand{\changed}[2]{#2}

\theoremstyle{thmstyleone}%
\theoremstyle{thmstyletwo}%

\theoremstyle{thmstylethree}%

\def \belletwo {Belle\,II\xspace}

\def \knn {$k$-NN\xspace}
\def \pnn {$p$-NN\xspace}
\def \enn {\textepsilon-NN\xspace}

\begin{document}

\title[Article Title]{Real-time Graph Building on FPGAs for Machine Learning Trigger Applications in Particle Physics}

\author*[1]{\fnm{Marc} \sur{Neu}}\email{marc.neu@kit.edu}
\author[1]{\fnm{Jürgen} \sur{Becker}}\email{juergen.becker@kit.edu}
\author[2]{\fnm{Philipp} \sur{Dorwarth}}\email{philipp.dorwarth@student.kit.edu}
\author[2]{\fnm{Torben} \sur{Ferber}}\email{torben.ferber@kit.edu}
\author[2]{\fnm{Lea} \sur{Reuter}}\email{lea.reuter@kit.edu}
\author[2]{\fnm{Slavomira} \sur{Stefkova}}\email{slavomira.stefkova@kit.edu}
\author[1]{\fnm{Kai} \sur{Unger}}\email{kai.unger@kit.edu}

\affil[1]{\orgdiv{Institut fuer Technik der Informationsverarbeitung (ITIV)}, \orgname{Karlsruhe Institute of Technology (KIT)}, \orgaddress{ \city{Karlsruhe}, \postcode{76131}, \country{Germany}}}
\affil[2]{\orgdiv{Institute of Experimental Particle Physics (ETP)}, \orgname{Karlsruhe Institute of Technology (KIT)}, \orgaddress{ \city{Karlsruhe}, \postcode{76131}, \country{Germany}}}

\abstract{We present a design methodology that enables the semi-automatic generation of a hardware-accelerated graph building architectures for locally constrained graphs based on formally described detector definitions. 
In addition, we define a similarity measure in order to compare our locally constrained graph building approaches with commonly used k-nearest neighbour building approaches.
To demonstrate the feasibility of our solution for particle physics applications, we implemented a real-time graph building approach in a case study for the \belletwo central drift chamber using Field-Programmable Gate Arrays (FPGAs).
Our presented solution adheres to all throughput and latency constraints currently present in the hardware-based trigger of the \belletwo experiment.
We achieve constant time complexity at the expense of linear space complexity and thus prove that our automated methodology generates online graph building designs suitable for a wide range of particle physics applications.
By enabling an hardware-accelerated pre-processing of graphs, we enable the deployment of novel Graph Neural Networks~(GNNs) in first level triggers of particle physics experiments.
}


\maketitle

\section{Introduction}\label{sec:intro}
Machine Learning is widely used in particle physics for various reconstruction tasks and Graph Neural Networks (GNNs) are recognised as one possible solution for irregular geometries in high energy
physics.
GNNs have proven suitable for jet clustering~\cite{Ju.2020},  calorimeter clustering~\cite{Wemmer:2023znh}, particle track reconstruction~\cite{DeZoort.2021, Duarte.2022, Ju.2020b}, particle tagging~\cite{Mikuni.2020, Qu.2020} and particle flow reconstruction~\cite{Pata.2021}.
However, all applications described above are implemented in an offline environment, relying on high performance computing clusters utilising Central Processing Units~(CPUs) and Graphics Processing Units~(GPUs) to achieve the required throughput for the analysis of collision events.
Therefore, existing implementations are not suitable for real-time particle tracking and reconstruction in trigger systems of particle detectors.

The realisation of GNNs on FPGAs for particle tracking is an active area of research~\cite{Duarte.2022, Elabd.2022, Qasim.2019, Iiyama.2020}.
Due to latency and throughput constraints, a suitable implementation meeting all requirements imposed by particle physics experiments is yet to be developed.
Especially the generation of input graphs under latency constraints is a challenge that has not received full attention so far in the evaluation of existing prototypes.
Current prototypes as described in~\cite{Elabd.2022,Duarte.2022} are trained on preprocessed graph datasets, taking into account geometric properties of detectors.
However, a holistic implementation of GNNs for triggers requires the consideration of the entire data flow chain.
This raises the question on how to build graphs under latency constraints in high-throughput particle physics applications.

In our work, we consider constraints from currently operating first level trigger systems~\cite{ATLAS:2020esi, Sirunyan.2020, Unger.2023b}: event processing rates in the order of \SIrange{10}{100}{\mega\hertz} and latencies in the order of \SIrange{1}{10}{\us} render the utilisation of compound platforms based on CPUs and Field Programmable Gate Arrays (FPGAs) used in other research areas infeasible~\cite{Liang.2021, Zhang.2021}. For example, the typical transmission latency between CPU and FPGA on the same chip is already larger than \SI{100}{\ns}, making up a considerable processing time~\cite{Karle.2022}.

To overcome the research gap, our work comprises the following contributions:
First, we outline existing nearest neighbour graph-building methods and evaluate their feasibility for trigger applications.
Second, we develop a methodology to transform formal graph-building approaches to hardware accelerated processing elements in an automated way.
Third, we evaluate our proposed toolchain on the \belletwo central drift chamber~(CDC), demonstrating the feasibility of our solution to build graphs under the constraints imposed by current trigger systems.

The paper is organised as follows: 
In \cref{sec:background}
we give an overview of related work on FPGA-accelerated graph building.
The CDC, the event simulation and details of the beam background simulation are described in \cref{sec:dataset}.
The methodology for transforming discrete sensor signals into a graphical representation is discussed in \cref{sec:graph}.
The procedure for implementing real-time graph building in hardware is described in \cref{sec:toolchain}.
A concrete example of real-time graph building for the Belle~II CDC is provided in \cref{sec:belleii_graph_building}.
We summarise our results in \cref{sec:conclusion}.

\FloatBarrier
\section{Related Work}\label{sec:background}
Previous work on FPGA-accelerated GNNs for particle physics utilise input graphs based on synchronous sampled collision events as input for training and inference of the respective networks~\cite{Duarte.2022, Thais.2022}.
Early studies made use of fully connected graphs which lead to scalability challenges for detectors with more than 10 individual sensors~\cite{Shlomi.2021}.
Typical particle physics trigger systems have much higher number of sensors though (see \cref{tab:trigger-input-comparison}).

\begin{table}[h]
    \caption{Input parameters for the first level trigger systems in three current particle physics detectors. 
    For CMS, \SI{95}{\percent} quantiles for the number of sensor hits per event is reported in~\cite{Elabd.2022}, while for the \belletwo~CDC~\cite{Abe.2010} and DUNE~\cite{Rossi.2022} the number of sensors inputs is given.}
    \label{tab:trigger-input-comparison}
    \centering
    \scriptsize
    \begin{tabular*}{\columnwidth}{@{\extracolsep{\fill}}rccc}
        \toprule
         & CMS
         & \belletwo
         & DUNE \\
         \cmidrule(lr){2-4}
         &  \cite{Elabd.2022, Hartmann.2020} 
         &  \cite{Abe.2010}
         &  \cite{Rossi.2022} \\
        \midrule
        Subsystem
        & Muon 
        & CDC
        & \makecell[c]{ProtoDune SP} \\
        \midrule
        \makecell[r]{Number of \\ Sensors}
        & \num{6500}
        & \num{14336}
        & \num{15360} \\
        \midrule
        \makecell[r]{Trigger Data \\ Input Rate}
        & \SI{40}{\mega\hertz}
        & \SI{32}{\mega\hertz}
        & \SI{2}{\mega\hertz} \\
        \bottomrule
    \end{tabular*}
\end{table}

Aiming to significantly reduce the maximum size of input graphs, the geometric arrangement of sensors in the detector has been considered recently~\cite{Ju.2020b, DeZoort.2021}.
Nevertheless, input graphs are currently generated offline, stored in the FPGA memory and are accessed over AXI\footnote{AXI: Advanced eXtensible Interface, is an on-chip communication bus protocol.}-Mapped Memory interfaces in prototype implementations~\cite{Elabd.2022}.
However, as sensors in detectors are read out as individual channels without providing relational information, the processing of input graphs must be considered as part of the critical path in online track reconstruction and trigger algorithms.  

While building suitable input graphs for neural networks is a rather recent application, general nearest neighbour (NN) graph building has been studied extensively in literature~\cite{Vaidya.1989, Callahan.1995, Connor.2008}.
In order to reduce the computational demand of NN graph-building algorithms, continuous efforts have been made towards building approximate graphs making use of local sensitive hashing~\cite{Gionis.1999, Hajebi.2011}, backtracking~\cite{Harwood.2016}, or small world graphs~\cite{Malkov.2020}.
Performance \changed{improvement}{improvements} from these algorithms have been demonstrated for applications targeting high-dimensional graphs containing more than $10^6$ vertices such as database queries~\cite{Besta.2021}.
There are two key challenges that limit the generalisation of these techniques in the particle physics trigger context.
First, $k$-nearest neighbour (\knn) algorithms inherently rely on sequential processing and present challenges in efficient parallelisation.
Second, while there is a wide range of graph-processing frameworks available (see Ref.~\cite{Gui.2019} for a survey on graph processing accelerators), none of them meet the stringent latency and throughput requirements of current particle physics trigger systems:
FFNG~\cite{Liu.2023} focuses on the domain of high-performance computing and therefore does not impose hard real-time constraints.
GraphGen~\cite{Nurvitadhi.2014} relies on external memory controllers which introduce additional latency into the system.
GraphACT~\cite{Zeng.2020,Zhang.2021} \changed{utilise}{utilises} preprocessing techniques on CPU-FPGA compound structures in order to optimise throughput and energy efficiency which again introduces non determinism and additional latency.
And lastly, current GNN accelerators like HyGCN~\cite{Yan.2020} or AWB-GCN~\cite{Geng.2020} use the previously described techniques to reduce the required system bandwidth and improve the energy efficiency of the inference.
They are therefore not suitable for particle physics applications.

\FloatBarrier
\section{Simulation and Dataset}\label{sec:dataset}
In this work, we use simulated \belletwo events to benchmark the graph-building algorithms.
The detector geometry and interactions of final state particles with the material are simulated using \texttt{GEANT4}~\cite{GEANT4:2002zbu}, which is combined with the simulation of a detector response in the Belle II Analysis Software Framework~\cite{Kuhr:2018lps}.
The \belletwo detector consists of several subdetectors arranged around the beam pipe in a cylindrical structure that is described in detail in Ref.~\cite{Kou:2018nap, Belle-II:2010dht}.
The solenoid’s central axis is the $z$-axis of the laboratory frame.
The longitudinal direction, the transverse $xy$ plane with azimuthal angle $\phi$, and the polar angle $\theta$ are defined with respect to the detector’s solenoidal axis in the direction of the
electron beam.
The CDC consists of 14336 sense wires surrounded by field wires which are arranged in nine so-called superlayers of two types: axial and stereo superlayers. 
The stereo superlayers are slightly angled, allowing for 3D reconstruction of the track.
In the simulated events, we only keep the detector response of the CDC.\\

We simulated two muons ($\mu^+$,$\mu^-$) per event with momentum \hbox{$0.5 < p < 5\,\text{GeV/c}$}, and direction \hbox{$17^{\circ} < \theta < 150^{\circ}$} and \hbox{$0^{\circ} < \phi < 360^{\circ}$} drawn randomly from independent uniform distributions in $p$, $\theta$, and $\phi$.
The generated polar angle range corresponds to the full CDC acceptance.
Each of the muons is displaced from the interaction point between 20~cm and 100~cm, where the displacement is drawn randomly from independent uniform distributions.\\

As part of the simulation, we overlay simulated beam background events corresponding to instantaneous luminosity of   \hbox{$\mathcal{L}_{\text{beam}}=6.5\times10^{35}$\,cm$^{-2}$s$^{-1}$}~\cite{Liptak:2021tog, Natochii:2022vcs}.
The conditions we simulate are similar to the conditions that we expect to occur when the design of the experiment reaches its ultimate luminosity.

\begin{figure}[ht]
\centering
\includegraphics[width=0.47\textwidth]{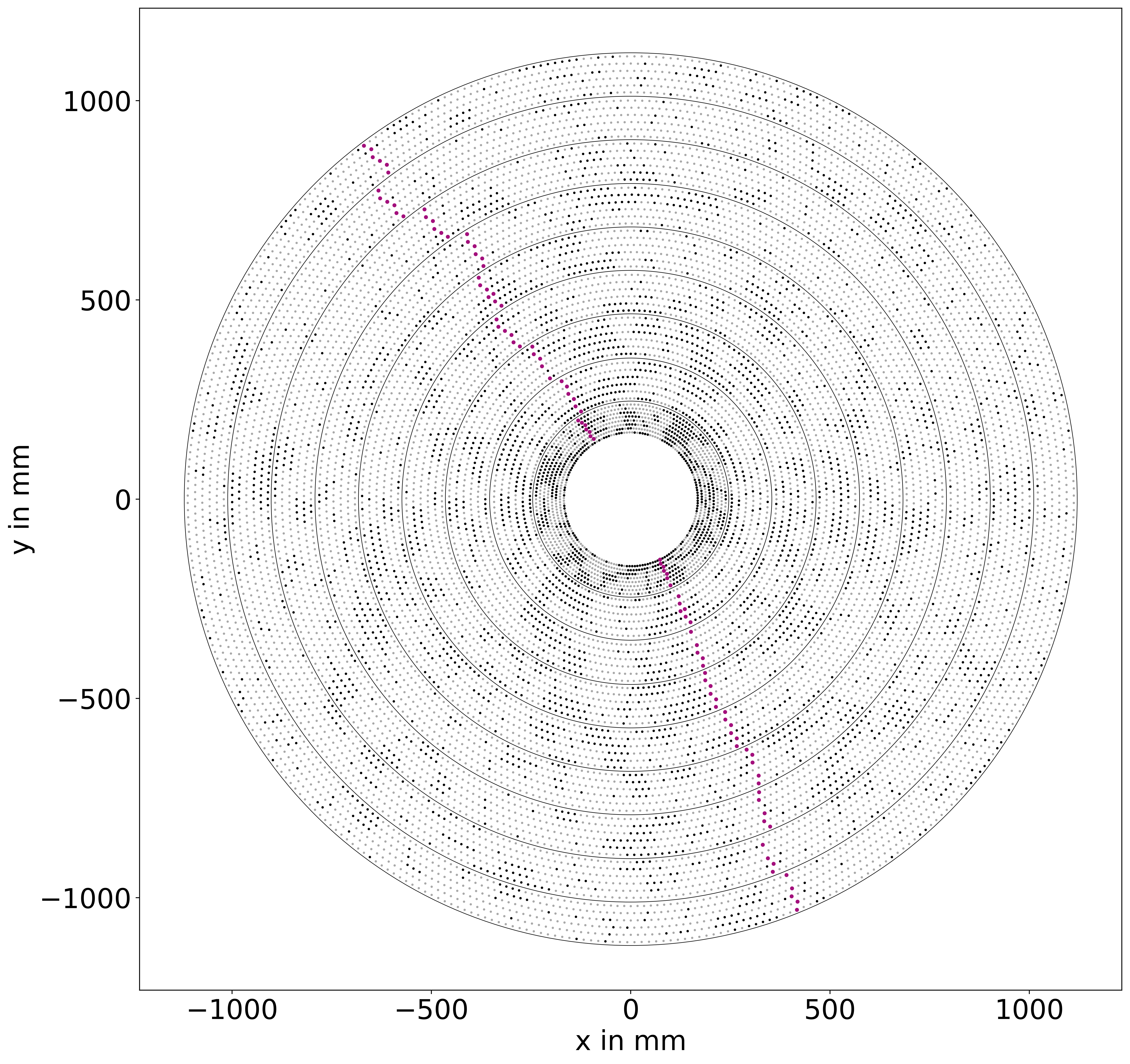}
\caption{Typical event display showing the transverse plane of the Belle~II CDC. Hits generated by signal muon particles are shown with purple markers and background hits by black markers.
} \label{fig:boringEventDisplay}\end{figure}

An example of an event display for a physical event $e^+e^-\to\mu^+\mu^-(\gamma)$ is shown in~\cref{fig:boringEventDisplay}. \changed{}{It is visible that the overall hit distribution of the exemplary event is dominated by the simulated beam background signal.}

\FloatBarrier
\section{Graph Building}\label{sec:graph}
This work proposes a methodology for transforming discrete sensor signals captured inside a particle detector into a graphical representation under real-time constraints.
Particular importance is given to the use-case of particle physics trigger algorithms, adhering to tight latency constraints in the sub-microsecond timescale.

Current large-scale particle detectors are composed of various discrete sensors and often, due to technical limitations, placed heterogeneously inside the system.
For this reason, signals from the sensors cannot be considered regularly distributed, as it is the case with, for example, monolithic image sensors.
In the following a detector $D$ is defined as a set of $N$ discrete sensors $\{ \Vec{s}_1, ... , \Vec{s}_N \}$, where each individual sensor $\Vec{s}_i$ is described by a feature vector of length $f$.
Some examples for described features are the euclidean location inside the detector, the timing information of the received signal, or a discrete \textit{hit identifier}. 
To map relational connections between individual sensors, a graph based on the detector description is generated which contains the respective sensor features.

Formally described, a graph building algorithm generates an non-directional graph $G(D,E)$, where $D$ is the set of vertices of the graph, and $E \subseteq D \times D $ is the set of edges.
The set of vertices is directly given by the previously described set of sensors in a detector.
Each edge $e_{ij} = e(\Vec{s}_i,\Vec{s}_j) \in E$ with $\Vec{s}_i, \Vec{s}_j \in D$ in the graph connects two sensors based on a building specification, that depends on sensor features.
In the following, we consider the case of building non-directed graphs.
We do not introduce any fundamental restrictions that limit the generalisation of our concept to directed graphs.

In general, graph building approaches are tailored to the specific detector and  physics case.
We consider three approaches that can be classified into two classes of nearest-neighbour graph building: locally constrained graphs, and locally unconstrained graphs.

\begin{figure}
     \centering
     \begin{subfigure}[b]{0.2\textwidth}
         \centering
         \includegraphics[width=\textwidth]{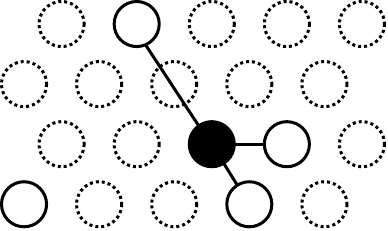}
         \caption{\knn, for $k=3$}
         \label{fig:kNN}
     \end{subfigure}
     \hfill
     \begin{subfigure}[b]{0.2\textwidth}
         \centering
         \includegraphics[width=\textwidth]{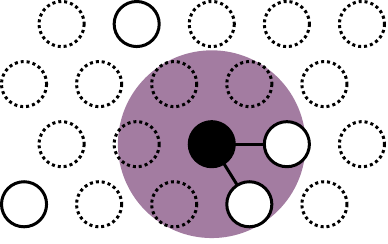}
         \caption{\enn}
         \label{fig:eNN}
     \end{subfigure}
    \hspace{0.25cm}
     \begin{subfigure}[b]{0.2\textwidth}
         \centering
         \includegraphics[width=\textwidth]{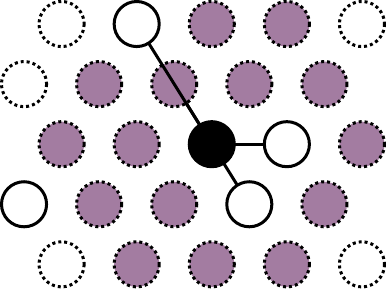}
         \caption{\pnn}
         \label{fig:patternNN}
    \end{subfigure}
        \caption{Example for the three different approaches of building nearest neighbour graphs. 
        Sensors inside a detector are depicted as circles. 
        A sensor which is hit by a particle is identified by a solid outline, those without a hit by a dotted outline. The query vertices are depicted in black. 
        Edges connecting two nearest neighbours are indicated by a solid line.
        Nodes filled with purple are considered candidate sensors, which are part of the specified search pattern around the query vertex.}
        \label{fig:two graphs}
\end{figure}

\Cref{fig:two graphs} depicts an exemplary cut-out of a detector, in which sensors are placed heterogeneously in two-dimensional space.
For simplicity, sensors are aligned in a grid-like structure without restricting the generality of our graph-building approach.
A graph is built for a query vertex which is depicted by a solid black circle.
We use the exemplary query vertex to illustrate NN-graph building on a single vertex for simplicity.
In the following, we compare the three building approaches and explain their differences.

\subsection{\knn}
\knn graph building is illustrated on a single query node in \cref{fig:kNN}.
Repeating the building algorithm sequentially leads to a worst-case execution time complexity of $\mathcal{O}(k \lvert D \rvert \log(\lvert D \rvert)$~\cite{Vaidya.1989}.
To reduce the execution time, parallelization of the algorithm has been studied in Ref.~\cite{Callahan.1995}, achieving a lower theoretical complexity.
Based on the optimization, a linear $\mathcal{O}(\lvert D \rvert)$ time complexity is achieved in experimental evaluation~\cite{Connor.2008}.
Nevertheless, substantial processing overhead  and limitations through exclusive-read, exclusive-write memory interfaces limit the usability for trigger applications.
To achieve a higher degree of parallelization, algorithms as described in Ref.~\cite{Hajebi.2011, Harwood.2016} make use of locally constrained approximate graphs.

\subsection{\enn}
\enn graph building is illustrated on a single query node in \cref{fig:eNN}.
The parameter \textepsilon~defines an upper bound for the distance of a candidate vertex from the query vertex.
All vertices for which \cref{eq:eNN} holds true are connected in a graph, yielding a locally constrained graph.
Figuratively, a uniform sphere is placed over a query point joining all edges which are inside the sphere into the graph:
\begin{equation}
    d(\Vec{x}_i,\Vec{x}_j) = {\lVert \Vec{x}_i - \Vec{x}_j \rVert}_2 < \epsilon
    \label{eq:eNN}
\end{equation}

Since the \enn approach is controlled by only one parameter, it is a general approach to building location-constrained graphs.
However, variations between adjacent sensors in heterogeneous detectors are not well represented in the \enn algorithm.

\subsection{\pnn}
Pattern nearest-neighbour (\pnn) graph building is illustrated on a single query node in \cref{fig:patternNN}.
For building the graph, every candidate sensor is checked and, if \changed{a}{the} predefined condition \changed{}{$p(\Vec{x}_i,\Vec{x}_j)$ in \cref{eq:pNN}} is fulfilled, the edge between candidate node and query node is included in the graph. \changed{}{ 
\begin{equation}
    p(\Vec{x}_i,\Vec{x}_j) \implies True 
    \label{eq:pNN}
\end{equation}
}

\subsection{Comparison}
When comparing the \knn, the \enn and the \pnn algorithms, it is obvious that in general all three approaches yield different graphs for the same input set of sensors.
\changed{
While the \pnn building and the \enn building can both be considered locally constrained algorithms,
}{
The \pnn building and the \enn building can both be considered locally constrained algorithms with differing degrees of freedom. While \enn building maps the locality into exactly one parameter, the definition of the \pnn building offers more flexibility. In contrast,}
the \knn approach differs as outliers far away from a query point might be included.
Nevertheless it is noted in Ref.~\cite{Prokhorenkova.2020}, that on a uniformly distributed dataset a suitable upper bound \textepsilon\textsuperscript{*} exists, for which the resulting \enn graph is a good approximation of corresponding \knn graph.

\FloatBarrier
\section{Toolchain}\label{sec:toolchain}
In the following, we leverage the described mathematical property to demonstrate the feasibility of building approximate \knn graphs for trigger applications.
First, we provide a methodology to evaluate the approximate equivalence of \knn, \enn and \pnn graph building approaches, providing a measure of generality for \knn parameters chosen in offline track reconstruction algorithms~\cite{DeZoort.2021, Rossi.2022}.
Second, we semi-automatically generate a generic hardware implementation for the \pnn graph building \changed{as an application-specific version of the \enn graph building}{}, thus demonstrating the feasibility of graph-based signal processing in low-level trigger systems.\changed{}{Since \enn graph building is a special case of \pnn graph building, we have also covered this case in our implementation.}
Third, we perform a case study on the \belletwo trigger system demonstrating achievable throughput and latency measures in the environment of trigger applications.

\subsection{Hardware Generator Methodology}
Algorithms that generate graphs by relating multiple signal channels belong to the domain of digital signal processing.
As such they share characteristics of typical signal processing applications like digital filters or neural networks.
Both applications are data-flow dominated and require a large number of multiply-and-accumulate operators and optimizations for data throughput.
Thus, implementing these algorithms on FPGAs \changed{show promising results}{improves latency and throughput} in comparison to an implementation on general purpose processors~\cite{Pfau.2018}.

\changed{
Developing custom digital logic for FPGAs is time-consuming and error-prone. 
To increase productivity, various high-level synthesis (HLS) frameworks have been developed that transform digital signal processing applications from formal definitions into hardware implementations, reducing the required design effort.
For example, digital filters are automatically implemented by commercially available development tools like MATLAB and hardware-aware training and deployment of neural networks is addressed by open-source tool-chains like FINN~\mbox{\cite{Umuroglu.2017, Blott.2018}} and HLS4ML~\mbox{\cite{Duarte.2018, FastMLTeam.2023}}.
While these frameworks have lowered the entry barriers for FPGA-algorithm development, their off-the-shelf usability is limited to pre-defined neural network architectures.
In addition, adapting the frameworks to support custom architectures is often time-consuming and error-prone.
}{
Various high-level synthesis (HLS) frameworks have been developed to reduce the required design effort such as FINN~{\cite{Umuroglu.2017, Blott.2018}} and HLS4ML~{\cite{Duarte.2018, FastMLTeam.2023}} with which the realisation of the GarNet, a specific GNN architecture, is possible.
Although these frameworks offer a low entry barrier for the development of FPGA algorithms, they are unsuitable for the implementation of our graph building concept.
}

\begin{figure}[ht]
\centering
\includegraphics[width=0.47\textwidth]{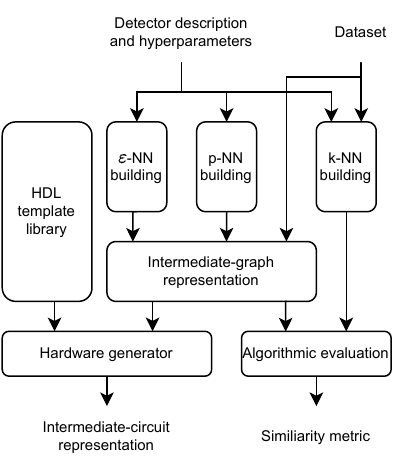}
\caption{Proposed generator-based methodology for our graph building approach. On the left side, the development flow for the hardware implementation is depicted, yielding an intermediate hardware representation. On the right side, flow for the algorithmic evaluation of the algorithms is shown.
}\label{fig:Methodology}
\end{figure}

\changed{Therefore, we propose a generator-based methodology enabling to transform a graph building algorithm into an actual firmware implementation.}{Therefore, we propose a generator-based methodology enabling to transform a graph building algorithm into an actual firmware implementation, that grants us complete design freedom at the register transfer level.}
\Cref{fig:Methodology} illustrates our development flow for both the generation of an intermediate representation of the circuit and an algorithmic evaluation of the building approach.
As an input a database containing the formal definition of a detector is expected alongside hyperparameters \changed{describing the building approach.}{, e.g. \textepsilon~ for the \enn graph building.}
Based on the selected approach, an intermediate-graph representation is generated, containing information\changed{}{on} how the building approach is mapped onto the detector.
The intermediate-graph representation serves as an input for the hardware generation and the algorithmic evaluation.

On one side, an intermediate-circuit representation
is generated by combining the intermediate-graph representation and parameterised hardware modules from our hardware description language~(HDL) template library.
\changed{}{The template library contains the elementary building blocks required to implement online graph building, in particular the static routing network, the edge processing elements and interface definitions.}
We use Chisel3~\cite{Bachrach.2012} as hardware-design language providing an entry point to register transfer-level circuit designs in Scala.

On the other side, the intermediate-graph representation is evaluated on a user-defined dataset and compared to a generic \knn graph-building approach.
To achieve a quantitative comparison we introduce similarity metrics for different operating conditions in the detector in \cref{sec:belleii_graph_building}.
This result can be used to iteratively adapt hyperparameters in the \enn or \pnn approach, improving the similarity to \knn graphs that are often used in offline track reconstruction.

\subsection{Intermediate-Graph Representation}
The parameter \textepsilon~ in the \enn approach and the pattern function in the \pnn approach limit the dimensionality of the graph under construction.
In comparison to fully-connected graphs, the maximum number of edges is lowered by imposing local constraints on the connectedness of sensors in the detector.
Local constraints are implemented by considering the influence of static sensor features, like euclidean distances between sensors, during design time of the FPGA firmware.
Leveraging the a-priori knowledge of the sensor position, the computational effort required during online inference of the algorithm is lowered.

\Cref{alg:staticGraph} describes the procedure to derive the intermediate-graph representation of an arbitrary graph-building procedure.
As an input the formally described set of sensors $D$ is given.
Iterating over every sensor in the detector, the locality of not yet visited sensors is checked by a user-defined \textit{metric} describing the graph building approach.
If a sensor is considered to be in the neighbourhood of another sensor, the connection is added to the resulting set of edge candidates $E$.
All edges in $E$ must be checked for their validity during the inference of the online graph building.

The combination of the formal detector description and the set of candidate edges is sufficient to describe an arbitrary building approach on non-directed graphs.
According to algorithm~\ref{alg:staticGraph}, the worst-case time complexity during design-time amounts to $\mathcal{O}({\lvert D \rvert }^2)$, which is higher than the worst-case time-complexity of state-of-the-art \knn building approaches.
However, the worst-case time-complexity during run-time is now only dependent on the number of identified edges during design-time.
Therefore, generating a graph of low dimensionality by choosing a suitable \textit{metric}\changed{,}{, e.g. a small \textepsilon~ in the \enn approach,} considerably lowers the number of required comparisons at run-time.
Such an optimization would not be possible when using a \knn approach, as even for a low dimensionality all possible edges must be considered.

\begin{algorithm}
\caption{Design-time graph building}\label{alg:staticGraph}
\hspace*{\algorithmicindent} \textbf{Input:} Set of Sensors $D$\\
\hspace*{\algorithmicindent} \textbf{Output:} Set of Edges $E$
\begin{algorithmic}[1]
\Procedure {buildGraph}{$D$}
\State $E \leftarrow \emptyset$
\While {$D \not\subset \emptyset$}
\State $s_i \leftarrow D.pop()$
\ForAll {$s_j \in D$}
\If { $metric(s_i,s_j)$}
\State $E \leftarrow E \cup \{e_{ij}\}$
\EndIf
\EndFor
\EndWhile
\State \textbf{return} $E$
\EndProcedure
\end{algorithmic}
\end{algorithm}

\subsection{Full Toolchain Integration}

Our methodology covers the conversion of an arbitrary graph building algorithm into an intermediate-circuit representation.
The resulting intermediate-circuit representation, implemented on the FPGA as a hardware module, exposes multiple interfaces on the FPGA.
On the input side, heterogeneous sensor data is supplied through a parallel interface as defined in the detector description.
On the output side, graph features are accessible through a parallel register interface to provide edge features to successive processing modules.

Considering the application of our module in a latency-sensitive, high-throughput environment like particle experiments, direct access to graph data is required at the hardware level.
Therefore bus architectures employed in general-purpose processors, like AXI or AMBA, are not suitable for our use case.
\changed{To reduce communication overhead between registers, which store graph data, and algorithmic processing units, an analysis of data paths during the generation of the final FPGA firmware is required.}{For this reason, our graph building module is connected to subsequent modules via buffered stream interfaces, reducing the routing overhead in the final design.}

\begin{figure}[!h]
\centering
\includegraphics[width=0.3\textwidth]{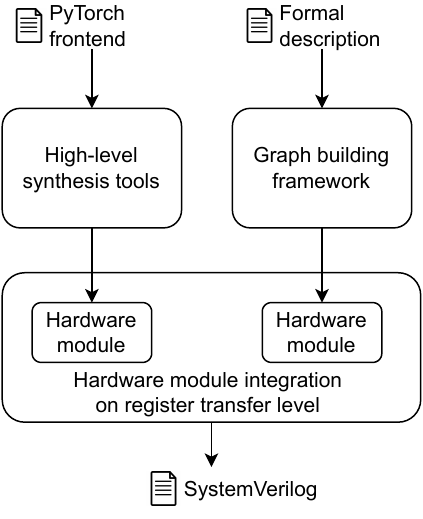}
\caption{Exemplary integration of our graph building methodology into a state-of-the-art HLS design flows.
}\label{fig:HLSFlow}
\end{figure}

\Cref{fig:HLSFlow} depicts exemplary, how our \changed{}{Chisel3-based} graph building methodology is combined with state-of-the-art HLS tools\changed{, enabling}{ such as HLS4ML~\cite{FastMLTeam.2023}, FINN~\cite{Umuroglu.2017, Blott.2018}
or ScaleHLS~\cite{Li.2021, Ye.2022} in order to enable} the generation
of hardware-accelerated neural networks.
The left side of the figure depicts a generic HLS flow converting, for example, a PyTorch~\cite{paszke2019pytorch} neural network model into hardware modules.
\changed{There are numerous HLS toolchains available for deploying neural networks on FPGAs, for example HLS4ML~\mbox{\cite{FastMLTeam.2023}}
, FINN~\mbox{\cite{Umuroglu.2017, Blott.2018}}
or ScaleHLS~\mbox{\cite{Li.2021, Ye.2022}}}{}.
The register transfer level description of hardware modules generated by HLS toolchains are
composed of discrete registers, wires, and synthesizable operations.
In a similar way, the right side of the figure depicts our proposed graph building procedure.
The formal detector description and the user-defined graph building \textit{metric} are used as an input to generate a register-transfer level description of the hardware module.
As both toolchains are generating hardware descriptions in the register transfer abstraction level, merging the two modules is feasible.
Last, a top level design combining both modules in SystemVerilog~\cite{SystemVerilog.2018} is generated for an FPGA-specific implementation using commercially available toolchains, for example Vivado~ML~\cite{Vivado.2022}.

\subsection{Module Architecture}

\begin{figure}[h]
\centering
\includegraphics[width=0.47\textwidth]{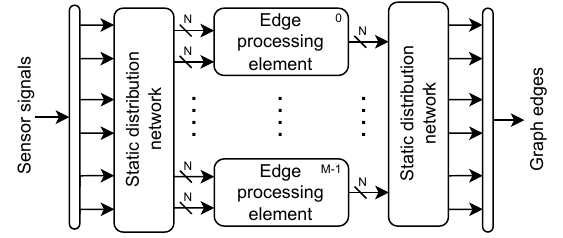}
\caption{System architecture of the generated hardware module. Sensor signals are received on the left side of the figure. The resulting graph edges are shown on the right side.}\label{fig:FirmwareOverview}
\end{figure}

Utilising the generated intermediate graph description, available generator templates, and user-defined hyperparameters, a hardware module is generated at the register-transfer level. 
The system architecture of the module is depicted in \cref{fig:FirmwareOverview}.
The total number of graph edges \changed{}{$\lvert E \rvert$} is factorised into $M$ edge processing elements and $N$ graph edges per edge processing element.
\changed{}{Time variant} readings from the detector sensors\changed{}{, e.g. energy or timing information,} are \changed{routed}{scattered} to an array of $M$ edge processing elements via a static distribution network.
\changed{}{In this way, each edge processing element has conflict-free access to the sensor data for classifying the respective edges.}
Every edge processing element builds $N$ graph edges in a time-division multiplex.
For each edge \changed{}{which is processed in an edge processing element}, data from two adjacent \changed{vertices}{sensors} are required which are provided to the edge processing element\changed{in two arrays of length $N$}{}.
\changed{}{Therefore, to process $N$ edges data from $2N$ sensors is required.}
Consequently, graph edges are built from candidates identified at design time yielding a sparse array of both active and inactive edges.
In the described architecture, all generated edges are accessible through parallel registers.
In case a serial interface is required for successive algorithms, an interface transformation is achieved by adding FIFO modules.

\begin{figure}[ht]
\centering
\includegraphics[width=0.47\textwidth]{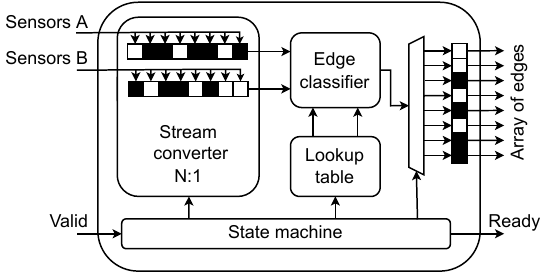}
\caption{The edge processing element consists of a stream converter, an edge classifier, and a lookup table.
Edge registers are made available through a parallel interface.}\label{fig:FirmwareEdge}
\end{figure}

\Cref{fig:FirmwareEdge} illustrates the block level diagram of an edge processing element in detail.
During design-time, each hardware module is allocated $N$ edges which are built sequentially.
Static allocation allows a-priori known sensor and edge features, like euclidean distances, to be stored in read-only registers.
During run-time, the described module loads static features from the registers, combines them with variable input features, like the deposited energy,
and classifies the edge as active or inactive.
The online graph building is carried out in three steps.
First, a pair of sensor readings is loaded from the shift registers, and static sensor and edge features are loaded from a static lookup table.
Second, a Boolean flag is generated based on a neighbourhood condition e.g., a user-specified metric is fulfilled for two adjacent sensors.
Third, the resulting feature vector of the edge is stored in the respective register.
Feature vectors of all edge processing elements are routed via a second static distribution network mapping each edge to a fixed position in the output register.

The proposed architecture takes advantages of distributed lookup tables and registers on the FPGA in two ways.
First, due to the independence of the edge processing elements space-domain multiplexing is feasible on the FPGA even for large graphs.
Second, static features of the graph edges and vertices are stored in distributed registers allowing logic minimisation algorithms to reduce the required memory~\cite{Harbaum.2016}.

To conclude, we developed an architecture for online graph building which is well suited for the latency constrained environment of low level trigger systems in particle physics experiments.
The variable output interface allows for an easy integration of successive trigger algorithms and leaves ample room for application specific optimisation.
The number of output queues is controlled by the parameter $N$ which yields a flexible and efficient design supporting variable degrees of time-domain multiplexing.

\FloatBarrier
\section{Case Study: Belle II Trigger}\label{sec:belleii_graph_building}
To demonstrate the working principle of our concept, we adapt our graph building methodology for the first level~(L1) trigger of the \belletwo experiment.
The implementation focuses on the CDC (see \cref{sec:dataset}) that is responsible for all track-based triggers.

\subsection{Environment}
The aim of the trigger system is to preselect collision events based on their reconstructed event topologies.
In order to filter events, a multi-stage trigger system is employed.
As a result, the effective data rate and thus the processing load of the data acquisition systems is reduced.
\begin{figure}[ht]
\centering
\includegraphics[width=0.47\textwidth]{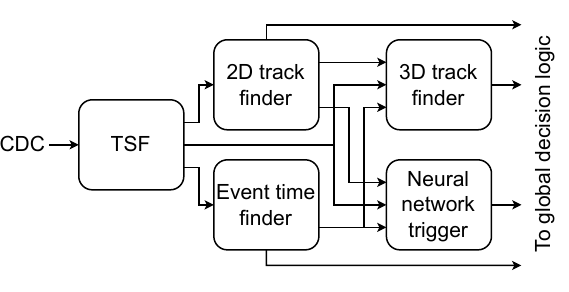}
\caption{Flowchart of the L1 trigger system at the \belletwo experiment, limited to systems that use the wire hit information from the CDC~\cite{Lai.2018}.}
\label{fig:cdc_trigger}
\end{figure}

To give an overview of the constraints and requirements imposed by the experiment, the existing system is briefly described in the following.
The L1 track triggers are shown schematically in in \cref{fig:cdc_trigger}.
They perform real-time filtering with a strict latency requirement of \SI{5}{\us}~\cite{Abe.2010}.
The sense wires inside the CDC are sampled with \SI{32}{\mega\hertz} and wire hits are accumulated for approximately \SI{500}{\ns}.
In order to process all available input signals concurrently, a distributed FPGA-based platform is employed.

To obtain a trigger decision, track segments are generated from incoming events in parallel by performing space-division multiplexing.
Based on the output of the track segment finder~(TSF), multiple algorithms including conventional 2D and 3D track finding algorithms as well as a Neural Network Trigger~\cite{Unger.2023b}
generate track objects of varying precision, efficiency, and purity for a Global Decision Logic~\cite{Lai.2018}.

The integration of GNNs in the L1 trigger system requires an online-graph building approach that is optimised for both latency and throughput.
In this case study, we employ our proposed toolchain to generate an application-specific graph-building module as described in the previous section while adhering to constraints in the challenging environment of the \belletwo experiment.

\subsection{Graph Building}

The wire configuration of the CDC is mapped onto the formal  detector definition from \cref{sec:graph}, using wires as discrete sensors.
These sensors are called nodes or vertices in the following.
Inside the L1 trigger system, three signals are received per wire: a \textit{hit identifier}, the \textit{TDC readout} and the \textit{ADC readout}, where TDC is the output of a time-to-digital converter measuring the drift time , and ADC is the output of an analogue-to-digital converter measuring the signal height that is proportional to the energy deposition in a drift cell.
Cartesian coordinates of the wires inside the detector are known during design time and used as static sensor features.
Additionally, the distance between two vertices, which is also known during design-time, is considered as an edge feature.

\begin{figure}[ht]
\centering
\includegraphics[width=0.47\textwidth]{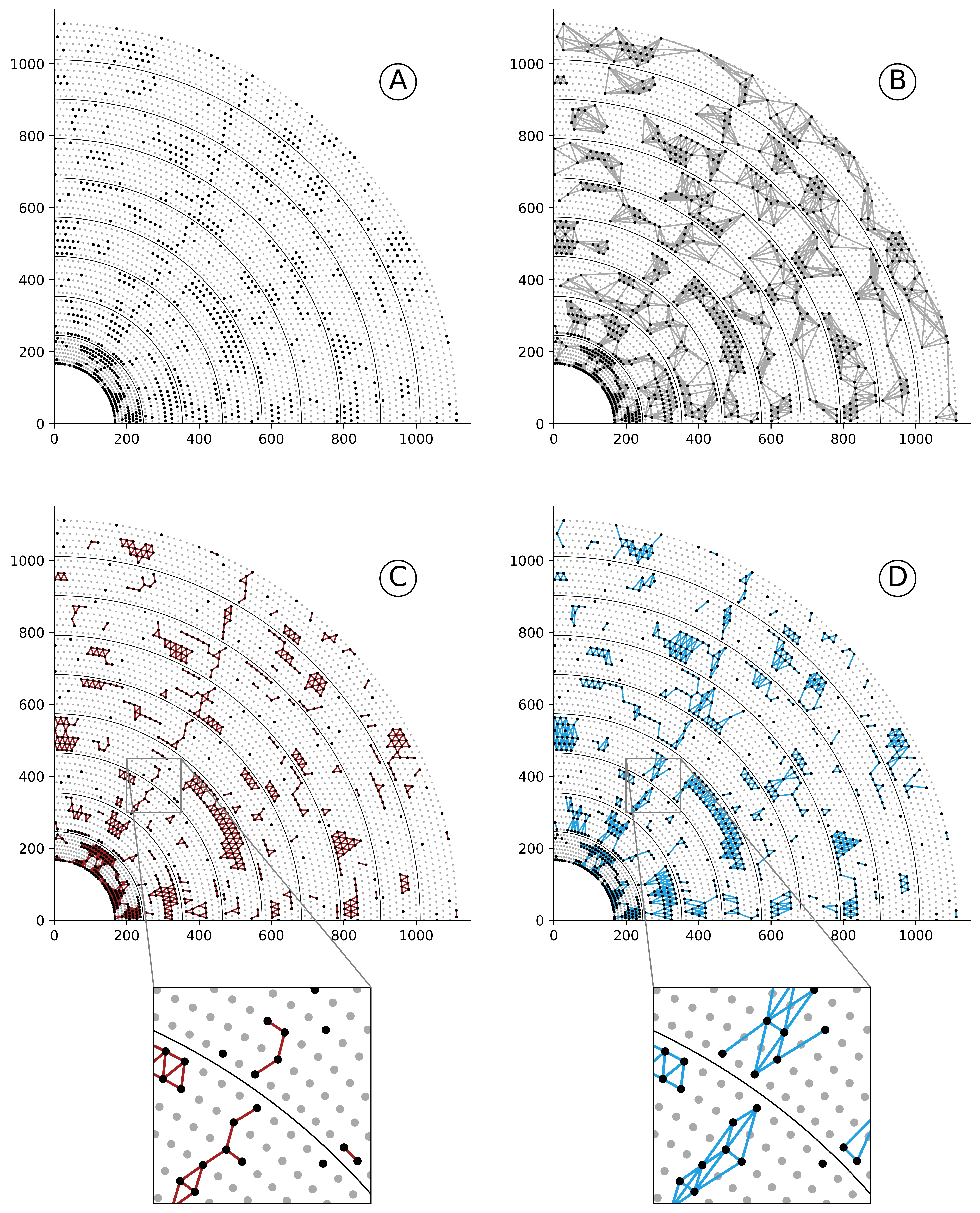}
\caption{Typical event display of the CDC for various graph building approaches. Quadrants show \Circled{A}~all hits, \Circled{B}~\knn graph building ($k$=6), \Circled{C}~\enn graph building (\textepsilon=\SI{22}{\mm}), and \Circled{D}~\pnn graph building (see \cref{fig:prototype:pattern}).
The inserts show zooms to a smaller section of the CDC.}\label{fig:ExemplaryEventDisplay}
\end{figure}

\begin{figure}[ht]
    \centering
    \includegraphics[width=0.4\textwidth]{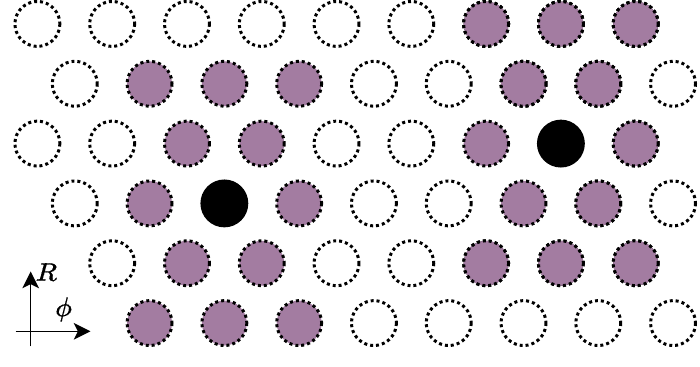}
    \caption{Two query vertices illustrate the neighbourhood pattern in hourglass shape used for the \belletwo detector case study. The superlayer is rolled off radially and an exemplary cut-out is shown. 
    Vertices which are considered neighbour candidates of the respective query vertex are shown as purple-filled markers.}
    \label{fig:prototype:pattern}
\end{figure}

Illustrating the working principle our graph building approaches, \cref{fig:ExemplaryEventDisplay} depicts four 
cut-outs of the CDC in the $x$-$y$ plane for $z=0$.\\
In sector \Circled{A}, \textit{hit identifier} received by the detector for an exemplary event are indicated by black markers.
The other three sectors show one graph building approach each:
Sector \Circled{B} depicts a \knn graph for of $k=6$, as there are up to six direct neighbours for each wire.
The \knn graphs connects wires that are widely separated.
Sector \Circled{C} shows an \enn graph for \textepsilon~=~\SI{22}{\mm}.
The specific value for \textepsilon~is chosen, because \SI{22}{\mm} is in the range of one to two neighbour wires inside the CDC. This graph building approach connects hits in close proximity only, yielding multiple separated graphs.
In addition, more edges are detected in the inner rings compared to the outer rings of the detector due to the higher wire density in this region.
Finally, sector \Circled{D} shows a \pnn graph using the pattern described in \cref{fig:prototype:pattern}.
The pattern extends the existing pattern~\cite{Pohl.2018, Unger.2020, Unger.2023} of the currently implemented TSF in the L1 trigger system by taking neighbours in the same superlayers into account.  
When comparing the \enn graphs and the \pnn graphs with each other, it is observed that the degrees\footnote{The degree of a vertex of a graph is the number of edges that are connected to the vertex.} of \pnn vertices are more evenly distributed (see inserts in \cref{fig:ExemplaryEventDisplay}).

\subsection{Parameter Exploration}

In general, \knn, \enn and \pnn algorithms generate different graphs for an identical input event.
However, to replace \knn graph building with a locally constrained graph building approach, the graphs should ideally be identical.
As the generated graphs \changed{depends}{depend} strongly on the chosen hyperparameters, on the geometry of the detector, and on the \changed{hit}{background} distribution of the events under observation, a quantitative measure of the similarity of the generated graphs between \knn graphs and locally constrained graphs, such as \enn or \pnn graphs, is necessary.
The optimal choice of the hyperparameter \textepsilon\textsuperscript{*} is the one that maximises the similarity for any $k$.
For this optimisation we use simulated events as described in \cref{sec:dataset}.
We generate both the \knn graphs and the locally constrained graphs  on the dataset considering the neighbourhood of wires inside the detector.
Edges of the \knn graphs are labelled $E_k$, whereas the edges of observed locally constrained graphs are labelled $E_l$.
We measure the similarity between the two graphs using \changed{the}{} the binary classifications metrics recall and precision
defined as
\begin{equation}
    recall = \frac{\vert E_k \cap E_l \vert}{\vert E_k\vert},
    \label{eq:recall}
\end{equation}
\begin{equation}
    precision = \frac{\vert E_k \cap E_l \vert}{\vert E_l\vert}.
    \label{eq:precision}
\end{equation}

\begin{figure*}
     \centering
     \begin{subfigure}[bt]{0.48\textwidth}
         \centering
         \includegraphics[width=\textwidth]{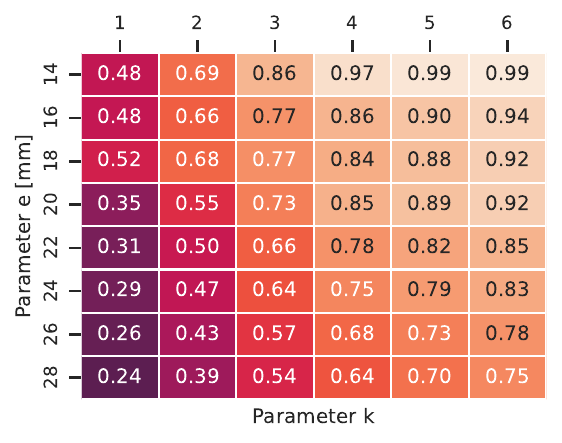}
         \caption{Precision}
         \label{fig:algorithm:precision}
     \end{subfigure}
     \hfill
     \begin{subfigure}[bt]{0.48\textwidth}
         \centering
         \includegraphics[width=\textwidth]{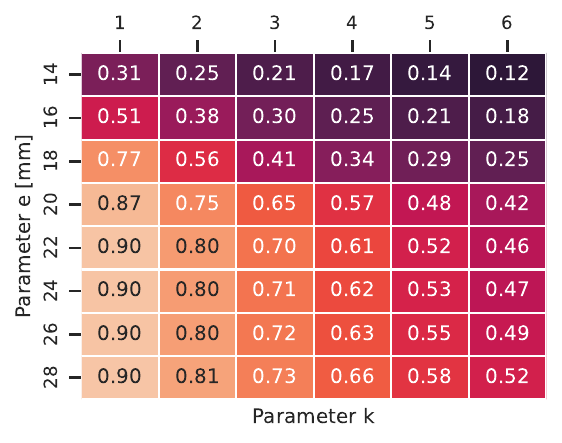}
         \caption{Recall}
     \end{subfigure}
        \caption{Precision and recall for the comparison of the \knn and \enn graph building approaches.}
        \label{fig:algorithm:results}
\end{figure*}

\changed{}{To perform the evaluation, we automate the parameter exploration using Python~3.10.}
We vary $k$  between \numrange{1}{6} and \textepsilon~ between \SIrange{14}{28}{\mm}, as the minimal distance between two wires in the CDC is approximately \SI{10}{\mm}.
Precision and recall scores are calculated for every pair of $k$ and \textepsilon~parameters and show mean value over \num{2000} events in \cref{fig:algorithm:results}.
As expected, the precision score increases monotonically when parameter $k$ is increased.
In addition, it increases if the parameter \textepsilon~is reduced.
The recall score behaves in the opposite way: It monotonically decreases when parameter $k$ is increased.
In addition, it decreases if the parameter \textepsilon~is decreased.
Similarity is defined as the ratio between recall and precision, where an optimal working point also maximizes reall and precision \changed{itself}{themselves}.
We observe that we do not find high similarity for all values of $k$.
Maximal similarity is found for $k=3$ and \textepsilon~$=\SI{22}{\mm}$, and  $k=4$ and \textepsilon~$=\SI{28}{\mm}$, respectively.
The corresponding precision and recall on the underlying data set are around 65-70\%.

The similarity between \knn and \enn graphs can be interpreted in relation to the mathematical statement from Ref.~\cite{Prokhorenkova.2020} (compare \cref{sec:graph}).
Based on the background noise and the large number of hits per \changed{events}{event}, we assume that the \textit{hit identifiers} in the dataset are approximately uniformly distributed.
Therefore, we expect that pairs of \knn and \enn graphs exist that exhibit a high degree of similarity, e.g. precision and recall scores close to one.
Our expectation is only partially met as the trade-off point reaches only about 65-70~\%.
\changed{One possible reason for the remaining difference between the two graphs is the underlying background noise.
Although the events are clearly dominated by noise, the influence on the hit distribution is not strong enough for higher similarity scores.}{The achieved metrics indicate, that the \knn graph-building approach from high level trigger algorithms may be replaced by the \enn graph-building approach in the first-level trigger and behave qualitatively similar.}

\begin{figure}[bt]
     \centering
     \includegraphics[width=0.4\textwidth]{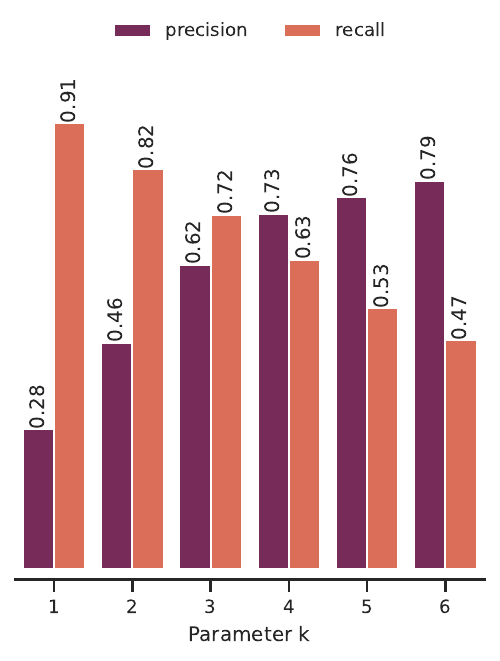}
     \caption{Precision and reall for the comparison between the \pnn graphs (for the pattern see in \cref{fig:prototype:pattern}) and the \knn graphs.
     }
     \label{fig:prototype:result}
 \end{figure}

We perform the same comparison between the \knn and the \pnn graph building approach as shown in \cref{fig:prototype:result}.
We achieve similar results in comparison to the \enn comparison: The recall score is monotonically decreasing for a larger parameter $k$, and the precision score is monotonically increasing for larger parameter $k$.
For $k$ between three and four, precision and recall scores are approximately similar and around 70~\%.

Again, our expectation of a high degree of similarity is only partially met.
This similarity is to be expected, as the chosen pattern is also locally constrained and approximately ellipsoid.

\subsection{Prototype Setup}

For the implementation of the proposed algorithm into a hardware prototype, the CDC is partitioned into 20 partially \changed{overlapped}{overlapped, independent} sectors in $\phi$ and radial distance $r$ for the L1 trigger.
Each \hbox{$\phi$-$r$-sector} is processed \changed{independently}{physically isolated} by one FPGA platform, the overlapping of the sectors ensures that no data is lost.
The overlapping sectors must be merged in subsequent reconstruction steps that are not part of the graph-building stage.
In the following, the graph-building module is implemented on the Belle~II Universal~Trigger~Board~4~(UT4) featuring a Xilinx~Ultrascale~XCVU160WE-2E.
The UT4 board is currently used in the \belletwo L1~Trigger and therefore serves as a reference for for future upgrades of the L1~trigger system.

To implement the online graph building module, we generate JSON~databases for every $\phi$-sector of the CDC.
Each database represents a formal detector containing the positions of the wires and information about sensor-features as described in section \cref{sec:graph}.
Sensor features are composed of \SI{1}{bit} for the binary \textit{hit identifier}, \SI{5}{bit} for the \textit{TDC readout}, \SI{4}{bit} for the \textit{ADC readout}, and the Cartesian coordinates of the wires.
Additional edge features containing information about the wire distances of two adjacent vertices are included as well.
The resolution of the euclidean features can be arbitrarily chosen and is therefore considered a hyperparameter of the module implementation.

The sector database and a function describing the pattern as illustrated in \cref{fig:prototype:pattern} is provided as an input to our proposed toolchain which is implemented in Python~3.10.
An intermediate graph representation is generated as a JSON~database, containing a type definitions of all vertices, edges and their respective features.
In addition, features known at design-time, such as Cartesian coordinates, are rounded down, quantized equally spaced, and included in the intermediate graph representation. 
By generating the databases for all 20 sectors, we identify the smallest and largest sector of the CDC to provide a lower and an upper bound for our problem size.
The maximum number of edges in each sector is determined by the pattern from \cref{fig:prototype:pattern}.
The smallest sectors are located in superlayer two containing \num{498} vertices and \num{2305} edges, while the largest sectors are located in superlayer six containing \num{978} vertices and \num{4545} edges.

To demonstrate our graph building approach, we synthesise the previously generated intermediate graph representation into a hardware module targeting the architecture of the UT4.
We provide the JSON~database as an input for the hardware generator, which is a set of custom modules implemented in Chisel~3.6.0.
In addition, we provide a Scala function that performs the online classification of edge candidates based on the \textit{hit identifier}:
an edge candidate is considered valid, if the \textit{hit identifiers} of both adjacent vertices are hit.
For the edge processing elements we choose the number of edges per edge processing element $N$ of eight.
Therefore, eight edges are processed sequentially in every edge processing element as described in \cref{sec:toolchain}.
Based on the required throughput of \SI{32}{\mega\hertz}, a system frequency of at least \SI{256}{\mega\hertz} is required to achieve the desired throughput.
By starting the generator application, edges and features are extracted from the intermediate graph representation and scheduled on edge processing elements.
After completion, the hardware generator produces a SystemVerilog file containing the graph-building hardware module~\cite{SystemVerilog.2018}.

\subsection{Implementation Results}

For further evaluation, the SystemVerilog module implementing the presented \pnn graph building is synthesised out-of-context for the UT4 board using Xilinx~Vivado~2022.2.
During synthesis, the target frequency $f_{sys}$ is set to \SI{256}{\mega\hertz}, for which no timing violations are reported by the tool.
In addition, functional tests are performed to validate the algorithmic correctness of the module.
In the following we perform two series of measurements to validate the feasibility of the proposed implementation on the Xilinx~Ultrascale~XCVU160WE-2E~FPGA.

\begin{figure*}[ht]
     \centering
     \begin{subfigure}[t]{0.48\textwidth}
         \centering
         \includegraphics[width=\textwidth]{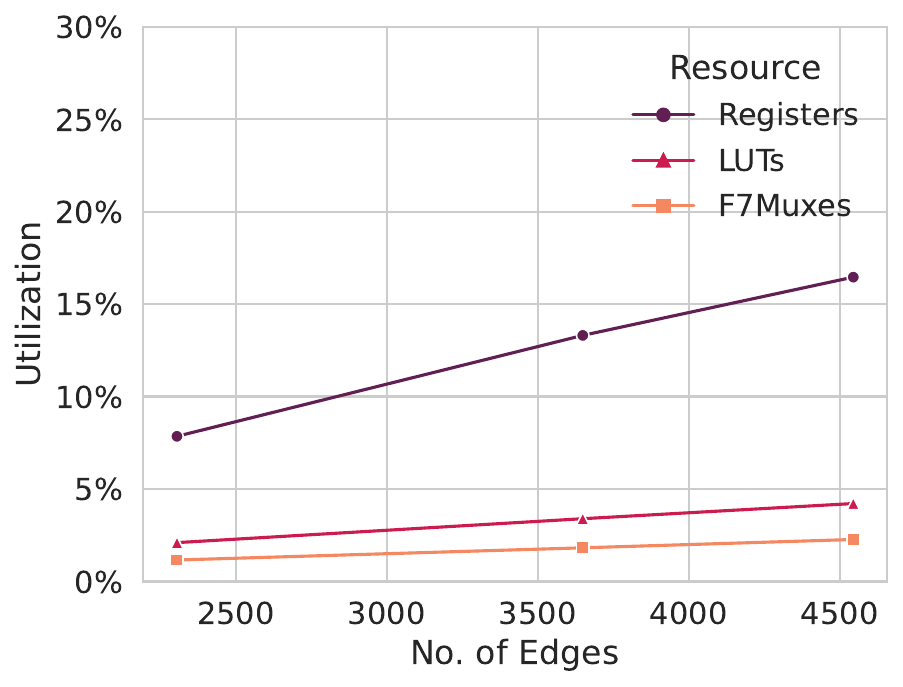}
         \caption{Utilization for a variable graph size $\lvert E \rvert$. The queue length parameter is set to eight, each edge is composed of \SI{60}{bits}.}
         \label{fig:implementation:graphsize}
     \end{subfigure}
     \hfill
     \begin{subfigure}[t]{0.48\textwidth}
         \centering
         \includegraphics[width=\textwidth]{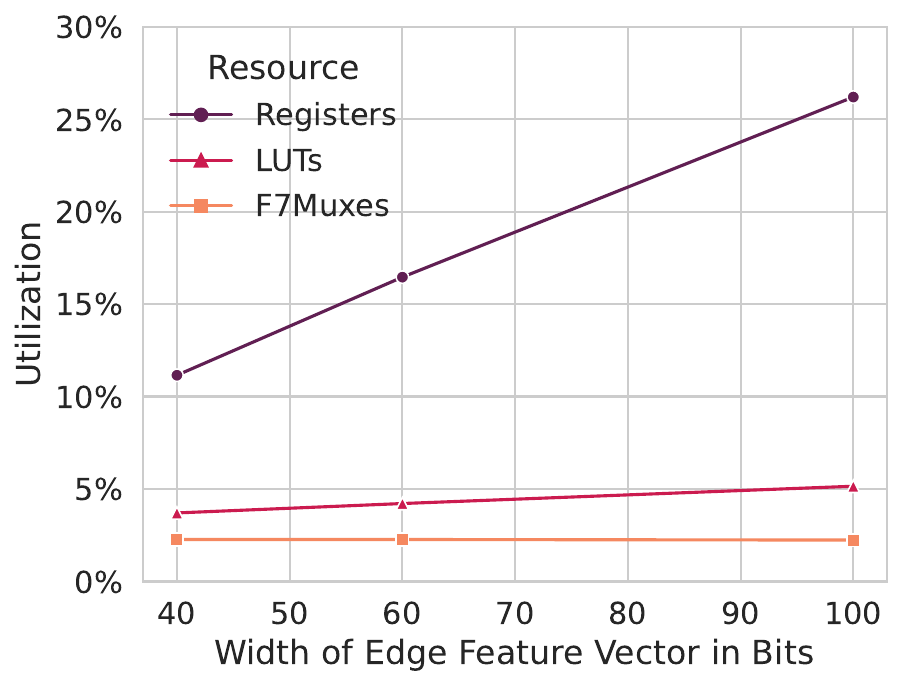}
         \caption{Utilisation for a variable edge width. The queue length parameter is set to eight, the input graph is composed of \num{4545} edges.}
         \label{fig:implementation:edgewidth}
     \end{subfigure}
        \caption{Resource utilisation reported after out-of-context synthesis on the UT4 platform using Vivado~2022.2 for registers, lookup tables~(LUTs) and multiplexers~(F7MUXes). Measurement are indicated by dots and connected by lines through linear interpolation to guide the eye. Unreported resource types are not utilised in the implementation.}
        \label{fig:implementation:result}
\end{figure*}

\Cref{fig:implementation:result} depicts the results of the two evaluation series, reporting the utilisation on the UT4 board for the respective resource types.
The first series of three synthesised versions is shown in \cref{fig:implementation:graphsize}, varying the input graph size in a suitable range between the \num{2305} and \num{4545} edges.
The highest occupancy is reported for registers, amounting up to \SI{16.46}{\percent} for the largest input graph, as opposed to \SI{7.84}{\percent} for the smallest graph.
For all other resource types, the utilisation is lower than \SI{5}{\percent}.
In general, it is observed that the resource utilisation scales linearly with the number of edges in the input graph.

\begin{table}[h]
    \caption{Overview of the features of the sensors used to define the edges. The occurrence indicates how often the respective feature is represented in an edge.}
    \label{tab:graph-edge-width}
    \centering
    {
    \begin{tabular*}{\columnwidth}{@{\extracolsep{\fill}}rccc}
        \toprule
        Feature
        & Type
        & Occurrence
        & Width\\
        \midrule
        \textit{hit identifier} 
        & Dynamic
        & 2
        & \SI{1}{bit} \\
        \textit{ADC readout} 
        & Dynamic
        & 2
        & \SI{4}{bit} \\
        \textit{TDC readout} 
        & Dynamic 
        & 2
        & \SI{5}{bit} \\
        \textit{X coordinate} 
        & Static
        & 2
        & \SIrange{4}{16}{bit} \\
        \textit{Y coordinate} 
        & Static
        & 2
        & \SIrange{4}{16}{bit} \\
        \textit{distance} 
        & Static
        & 1
        & \SIrange{4}{16}{bit} \\
        \bottomrule
    \end{tabular*}}
\end{table}

For the second series, a variation in resolution of the underlying edge features is considered. 
An overview of all utilised features is given in \cref{tab:graph-edge-width}. 
The width of features that are received as inputs from the CDC, namely \textit{hit identifier}, \textit{ADC readout}, and \textit{TDC readout}, are exemplary chosen in a way which is supported by the current readout system.
As an example, the \changed{the}{} \textit{TDC readout} quantisation of \SI{5}{bit} derives from the drift time resolution of \SI{1}{\ns} at a trigger data input rate of \SI{32}{\mega\hertz}.
The resolution of euclidean coordinates and distances can be optimised at design-time.

In the following, we choose a resolution between \SIrange{4}{16}{bit} which results in a quantisation error for the euclidean coordinates in the range \SIrange{34.4}{0.017}{\mm}. 
\SI{4}{bit} per coordinate result in a total edge width of \SI{40}{bit}, whereas a resolution of \SI{16}{bit} per coordinate results in a total edge width of \SI{100}{bit}.

The implementation utilisation of all three synthesised modules is shown in \cref{fig:implementation:edgewidth}, varying the resolution of euclidean coordinates and distances in the generated edges.

Similar to the previous measurement, the highest utilisation is reported for registers, taking up between \SI{11.1}{\percent} and \SI{26.1}{\percent} depending on the width of the edges.
\changed{However it}{It} can be seen, that the implementation size scales linearly with the \changed{number of edges in the input graph}{width of the graph edges. Increasing the resolution of a parameter, e.g. the \textit{TDC readout}, therefore leads to a proportionally higher utilisation of the corresponding resource on the FPGA}.

\begin{table*}[h]
    \caption{Utilization for variable graph size $\lvert E \rvert$, $\lvert V \rvert$, and edge width. Numerical implementation results are identical to the values shown in \cref{fig:implementation:result}.}
    \label{tab:implementation:result}
    \centering
    \scriptsize
    \begin{tabular*}{\textwidth}{@{\extracolsep{\fill}}ccccccccc}
        \toprule
         No. of Vertices  & No. of Edges &  Width of Edge & \multicolumn{2}{c}{{Registers}} & \multicolumn{2}{c}{{LUTs}} & \multicolumn{2}{c}{{F7Muxes}}\\
         \cmidrule(lr){4-5}\cmidrule(lr){6-7}\cmidrule(lr){8-9}
         &&& {abs.} & \si{\percent}  & {abs.} & \si{\percent}  & {abs.} & \si{\percent}\\
        \midrule
        498 & 2305 & {60} \si{bit} & \num{145333} & \SI{7.84}{\percent} & \num{19370} & \SI{2.09}{\percent} & \num{5760} & \SI{1.15}{\percent}\\
        786 & 3649 & {60} \si{bit} & \num{246511} & \SI{13.30}{\percent} & \num{31360} & \SI{3.39}{\percent} & \num{9120} & \SI{1.81}{\percent}\\
        978 & 4545 & {40} \si{bit} & \num{206573} & \SI{11.15}{\percent} & \num{34252} & \SI{3.70}{\percent} & \num{11360} & \SI{2.26}{\percent}\\
        978 & 4545 & {60} \si{bit} & \num{304919} & \SI{16.46}{\percent} & \num{38968} & \SI{4.21}{\percent} & \num{11360} & \SI{2.26}{\percent}\\
        978 & 4545 & {100} \si{bit} & \num{485473} & \SI{26.20}{\percent} & \num{47642} & \SI{5.14}{\percent} & \num{11200} & \SI{2.23}{\percent}\\
        \bottomrule
    \end{tabular*}
\end{table*}

Based on the presented results, the implementation of the graph building module is considered feasible on the UT4 board.
By experimental evaluation we show that our hardware architecture can be implemented semi-automatically for the L1 trigger of the \belletwo experiment, enabling the deployment of GNNs in the latency-constrained trigger chain.
The feature vectors of the edges are provided via a parallel output register, where the address of every edge is statically determined at design time.
Depending on successive filtering algorithms, any number of output queues can be provided.
To conclude, our toolchain allows for a flexible and resource efficient design of online graph building modules for trigger applications.
In the presented implementation, our module is able to achieve a throughput of \num{32} million samples per second at total latency of \SI{39.06}{\nano\second}, corresponding to ten clock cycles at $f_{sys}$.
As the reported latency is well below the required $\mathcal{O}(\SI{1}{\us})$, our graph building module leaves a large part of the latency and resource budget on FPGAs to the demanding GNN solutions.

\section{Conclusion}\label{sec:conclusion}
In our work, we analysed three graph building approaches on their feasibility for the real-time environment of particle physics machine-learning applications.
As the \knn algorithm, which is favoured by state-of-the-art GNN tracking solutions, is unsuitable for the strict sub-microsecond latency constraints imposed by trigger systems, we identify two locally constrained nearest neighbour algorithms \enn and \pnn as possible alternatives.
In an effort to reduce the number of design-iterations and time-consuming hardware debugging, we develop a generator-based hardware design methodology tailored specifically to online graph-building algorithms.
Our approach generalises graph-building algorithms into \changed{a}{an} intermediate-graph representation based on a formal detector description and user-specified metrics. 
The semi-automated workflow enables the generation of FPGA-accelerated hardware implementation of locally constrained nearest neighbour algorithms.
To demonstrate the capabilities of our toolchain, we perform a case study on the trigger system of the \belletwo detector.
We implement an online graph-building algorithm which adapts the pattern of the current track segment finder, demonstrating the feasibility of our approach in the environment of particle physics trigger applications.
The code used for this research is available open source under Ref.~\cite{sw_graphbuilding}.

Nearest neighbour algorithms presented in this work achieve a $\mathcal{O}(1)$ time complexity and a $\mathcal{O}(\lvert E \rvert)$ space complexity, compared to a $\mathcal{O}(\lvert D \rvert)$ time complexity in approximate \knn algorithms or a $\mathcal{O}(k \lvert D \rvert \log(\lvert D \rvert)$ complexity in the sequential case~\cite{Connor.2008, Vaidya.1989}.
As a result, our semi-automated methodology may also be applied to other detectors with heterogeneous sensor arrays to build graphs under latency constraints, enabling the integration of GNN-tracking solutions in particle physics.\\

During the evaluation of our similarity metric, we found a non-negligible difference between \knn graphs and locally constrained NN-graphs.
For the complete replacement of \knn graphs with our proposed \enn and \pnn graphs, the differences must be taken into account to achieve optimal performance when designing successive trigger stages.
For this reason, we consider the future development of methods for algorithm co-design essential for integrating GNNs into real-world trigger applications.
\changed{}{Careful studies of possible difference between simulated data are another main direction of future work.}

\backmatter

\bmhead{Data Availability Statement}%
The datasets generated during and analysed during the current study are property of the \belletwo collaboration and not publicly available.

\bmhead{Code Availability Statement}%
The code used for this research is available open source under Ref.~\cite{sw_graphbuilding}.

\bmhead{Acknowledgements}
The authors would like to thank the Belle~II collaboration for useful discussions and suggestions on how to improve this work.\\
It is a great pleasure to thank (in alphabetical order) Greta~Heine, Jan~Kieseler, Christian~Kiesling and Elia~Schmidt for discussions, and Tanja~Harbaum, Greta~Heine, Taichiro~Koga, Florian~Schade, and Jing-Ge~Shiu for feedback and comments on earlier versions of the manuscript.

\section*{Compliance with ethical standards}

\subsection*{Conflict of interest}
The  authors  declare  that  they  have  no  conflict  of  interest.

\clearpage
\bibliography{sn-bibliography}
\end{document}